\documentclass{article}
\usepackage[dvips]{graphicx} % Required for inserting images
\usepackage{float,bm,physics,slashed}
\usepackage{amsmath,amssymb,tabularx,booktabs}
\usepackage[titletoc]{appendix}
\usepackage[usestackEOL]{stackengine}
\usepackage{lipsum}
\edef\tmp{\the\baselineskip}
\setstackgap{L}{\tmp}
\usepackage{extarrows} 
\usepackage{color}
\usepackage[caption=false]{subfig}
\usepackage[colorlinks=true,linkcolor= blue,filecolor= blue,urlcolor= blue,citecolor= blue]{hyperref}
\usepackage{geometry} % Adjust page margins if needed
\usepackage{array,appendix}    % Required for specifying column widths
\begin{document}
\title{Sea-quark dynamics in decuplet ($\frac{3}{2}^+$) $\rightarrow$ octet ($\frac{1}{2}^+$) transition quadrupole moment}
\author{Preeti Bhall* and Alka Upadhyay}
\date{Department of Physics and Material Science, \\
Thapar Institute of Engineering and Technology, Patiala, Punjab-147004\\
\today}
\maketitle
\thanks{Email: preetibhall@gmail.com},
\thanks{alka.iisc@gmail.com}
\begin{abstract}
We investigated the electromagnetic quadrupole transition of baryon decuplet ($J^P= \frac{3}{2}^+$) to octet ($J^P= \frac{1}{2}^+$) using the statistical framework together with the principle of detailed balance. The statistical approach assumed the expansion of hadrons in terms of various quark-gluon Fock states. By specifying the appropriate multiplicity in spin, color $\&$ flavor space, the relative probabilities of strange and non-strange quark-gluon Fock state are calculated. These probabilities further accumulated in the form of statistical parameters, highlighting the importance of sea quarks and gluons in the electromagnetic transition. Our calculations includes the individual contribution of valence and sea (scalar, vector and tensor ) to the transition moment of baryons. The effect of flavor SU(3) symmetry and its breaking in both valence and sea quarks is studied by incorporating the strange quark mass. The strangeness in the sea is constrained by a suppression factor $(1-C_l)^{n-1}$, which depends upon the free energy of gluons. The computed results get affected upto 60 $\%$ and exhibit the dominance of octet sea. The present work has been compared with updated experimental data and various theoretical predictions. The results obtained may offer important insights for future experimental studies. %Moreover, the parameter ‘r’ for estimations of the role of the symmetry breaking in valence quark has been induced. The effect of this parameter is very small.
\end{abstract}
\section{Introduction}
Over the decades, significant extensive experiments are focused by various facilities such as JLab, MAMI, BESIII and HADES-GSI in order to comprehend the internal dynamics of hadrons. The study of electromagnetic properties of baryons (masses, spin distribution, magnetic moment, charge radii, quadrupole moment etc.) is one of the most interesting domain that reveals how the baryons are influenced by its constituents and also yields useful information about the underlying dynamics of quarks and gluons within the non-perturbative regime of QCD. Moreover, the scrutiny of the EM transition from baryon decuplet (spin -$\frac{3}{2}^+$) to octet (spin -$\frac{1}{2}^+$) also yield valuable information on the quark-quark interaction as well as the probable deformation in the structure of baryons. Considerable efforts have been made to investigate nuleon-to-$\Delta$(1232) resonance \cite{1,Perdrisat:2006hj,Punjabi:2015bba,Sparveris2013,Thoma:2007bm,PhysRevC.95.065205}, as $\Delta$(1232) is the lowest-energy excited state of N(939). Experimentally, the decay of $\Delta$- resonance into the nucleon and pion ($\Delta \rightarrow N \pi$) is the dominant branch, accounting for 99.4$\%$ of its total decay width, while the next significant decay channel is – electromagnetic, $\Delta\rightarrow N\gamma$, which contributes 0.6$\%$ \cite{PhysRevLett.79.4337}. Recently, the MicroBooNE collaboration has reported $\Delta$(1232) radiative decay through neutrino induced neutral current \cite{PhysRevLett.128.111801}. Following the spin- parity selection rule, the $\Delta^+ \rightarrow p\gamma$ transition amplitudes contain the magnetic dipole ($G_{M1}$), electric quadrupole ($G_{E2}$), and coulumb quadrupole ($G_{C2}$) contributions. The amplitudes $G_{M1}$, $G_{E2}$ and $G_{C2}$ give information about the magnetic dipole moment, electric quadrupole moment and charge quadrupole moment respectively \cite{PhysRevC.64.025203,PhysRevLett.93.212301,PhysRevLett.78.606}. At low momentum transfer, particularly in s-wave state, the transition M1 (magnetic dipole) mostly occurs because of the change in spin and isospin of a quark. Apart from this, at higher order, the d-state components with orbital angular momentum L=2 allow the electric quadrupole (E2) and Coulumb (C2) transition. These transitions manifest the deviation in nucleon and $\Delta$ from spherical symmetry. Accordingly, the amount of deformation can be quantified by the multipole ratios such as $R_{EM}$ =$\frac{E2}{M1}$ \cite{PhysRevLett.93.212301,Workman:2022ynf,Tiator2003} and $R_{CM}$= $\frac{C2}{M1}$ \cite{ PhysRevLett.86.2237,refId0,Abada:1995db}. 
% The study of $\Delta$ resonances have been focused through various theoreticaland phenomenological approaches like Isgur-Karl model [15] with modifiedrelativised approach [16], spontaneous chiral symmetry breaking through Goldstoneboson exchange[17, 18], QCD SUM Rules [20], light-frontmodel [21], Lattice QCD [23]. 
Presently, the spectrum of $\Delta$ resonance is not only limited to the high energy area but also in the field of astrophysics, where the existence of $\Delta$s is observed in neutron star \cite{Sahoo:2017flf}. The other transitions involving hyperons ($\Sigma^{*+}\rightarrow \Sigma^+, \Sigma^{*0}\rightarrow \Lambda, \Xi^{*-}\rightarrow \Xi^- $) are also crucial for gaining insight into the internal configuration of baryons \cite{PhysRevC.71.054609,BUTLER199369}.\\
%amplitudes between the $J^P =\frac{1}{2}^+$ nucleon and its $J^P =\frac{3}{2}^+$ resonance manifest the  
% The electromagnetic transition between the nucleon and excited baryons has been shown a very important method to study the structure of nucleon and baryon resonances [2]. % In particular, the very precise electron-scattering data from CLAS at Jefferson Lab (JLab) on light nucleon resonances had a leading role in the current progress [].In last few years, there have been numerous studies dealing with the transition properties of baryons from both theoretical and experimental point of view []. 
In the quark-model picture, the very first triumph was the estimation of $\Delta$(1232) $\rightarrow N\gamma$ transition magnetic moment \cite{PhysRevLett.13.514}. Also, the NRQM predicts the electric quadrupole transition moment $Q_{\Delta^{+} \rightarrow p}$ to be zero, which indicates that $\Delta$ isobar has a symmetrically spherical shape. But experimentally, it is found that the $\Delta \rightarrow$ N quadrupole transition is small and has some finite value \cite{PhysRevC.64.025203}. Both the LEGS \cite{PhysRevLett.79.4337} and the Mainz \cite{PhysRevLett.78.606} collaborations measured the quadrupole transition moment of $\Delta^{+}\rightarrow$ N yielding values of -0.018 $\pm$ 0.009 $\pm$ 0.034 fm$^{2}$ and -0.0846$\pm$ 0.0033 fm$^{2}$ respectively.
% There  is no experimental information available for the quadrupole moment of octet and decuplet baryons. Theoretically, various frameworks such as  relativistic quark model [], light-front holographic method [], QCD sum rule [], Lattice QCD [], chiral constituent quark model ($\chi$CQM) [], General Parameterization (GP) method [] examined the quadrupole moment of octet and decuplet baryons. On the other hand, the studies of EM quadrupole transition moment are rather limited.
% Since the decuplet baryons have very short lifetime, he quadrupole transition moment from decuplet to octet baryons are difficult to understand and very less known as decuplet baryons have very short lifetime.\\
The quest towards the understanding of internal dynamics of matter, EM form factors are the most essential quantities and directly related to the low-energy static properties of baryons. Recently, in 2022 \cite{10.1093/nsr/nwab187} the BESIII collaboration measured the form factors of baryons
within the time-like region. Several theoretical approaches such as the chiral perturbation theory \cite{PhysRevD.49.3459}, chiral effective theory \cite{PhysRevD.60.054022} Skyrme model \cite{doi:10.1142/S0217732395001137,transition}, QCD sum rules \cite{ALIEV2006291}, large $N_c$ limit \cite{Buchmann:2003fm,PhysRevLett.89.242001} and Lattice QCD \cite{PhysRevD.69.114506,PhysRevLett.94.021601,PhysRevD.77.085012} studied the EM tansition form factor and related properties of baryons. In Ref. \cite{PhysRevD.69.114506,PhysRevLett.94.021601}, with the lattice QCD simulation, N to $\Delta$ transition form factor was computed. In Ref. \cite{PhysRevD.77.085012}, the electromagnetic $N\rightarrow  \Delta$ transition FF's are calculated using two degenerate flavors of dynamical Wilson fermions as well as dynamical staggered fermions considering sea along with constituent quarks. The EM transition form factors and $\Sigma^0 \rightarrow \Lambda$ transition is studied in space-like momentum region within the framework of both Dyson Schwinger and covariant BSE \cite{Sanchis-Alepuz2018}.\\
% The electromagnetic form factors of the $\Delta(1232) \rightarrow N\gamma$ transition were evaluated both in quenched lattice QCD and using two dynamical Wilson fermions. 
 The electromagnetic structure of baryons is complex and affected by both valence and gluonic degrees of freedom in terms of the “sea". The "sea" is assumed to be dynamic and is filled with an infinite no. of $q\bar q$ pairs and gluons. The impact of strange ($s\bar s$) and non-strange sea ($u\bar u, d\bar d$) on the electromagnetic properties has been confirmed by various experimental collaborations like STAR \cite{PhysRevD.99.051102}, NuTeV \cite{ALBERICO1997471,Bazarko1995}, $\bar P$ANDA-GSI \cite{Barucca2021}, J-PARC \cite{aoki2021extension}. In 2019, the STAR experiment detected the prominent contribution of sea antiquarks to proton's spin. Additionally, NuTeV collaboration calculated the ratio $\frac{2(s+ \bar s)}{u+ \bar u+ d+ \bar d}$ = 0.477$\pm$0.063$\pm$ 0.053 \cite{Bazarko1995} which confirmed the 'strangeness' contribution to overall nucleon momentum. The quadrupole transition moment of baryons also get influenced by various factors, including sea quarks, spin and angular momentum contribution, confinement on quark masses etc. In literature, N. Sharma and H. Dahiya \cite{Dahiya_2018,PhysRevD.81.073001,SHARMA2013} examined the different transition moments from the $J^P =\frac{3}{2}^+$ decuplet to $J^P =\frac{1}{2}^+$ octet baryons using the chiral constituent quark model ($\chi$CQM). The individual effect of valence and sea quarks on different electromagnetic properties is calculated. In ref. \cite{PhysRevD.103.074025}, the EM transition of singly charmed baryons spin- $\frac{3}{2}^+$ studied with valence and sea contribution using pion mean-field approach. A.J. Buchmann and E.M. Henley \cite{PhysRevLett.93.212301,Buchmann2018,PhysRevD.65.073017,GeorgWagner_2000} employed the GP method to calculate the quadrupole transition moments of baryons. In ref. \cite{PhysRevD.49.3459}, the heavy baryon chiral perturbation theory framework calculates the decuplet to octet transition magnetic moments and electric quadruple moments of higher order. In Ref.\cite{PhysRevD.80.013008}, the authors analyzed the valence quark contributions to the $\Delta(1232) \rightarrow N\gamma$ transition in the lattice QCD regime using the covariant spectator approach. The radiative decays of decuplet to octet baryons are examined by light cone QCD sum rules \cite{ALIEV2006291}. Despite much theoretical work, the information on the quadrupole transition moment, especially accounting for the effect of sea quarks and gluons, is very limited. \\
The statistical model with the principle of detailed balance is a very useful framework to understand the hadronic structure, taking into account the individual contribution of valence and sea quarks. The model has been successfully applied to calculate various static properties like masses \cite{Kaur20166}, magnetic moments \cite{10.1093/ptep/ptx068,Kaur20167}, charge radii and quadrupole moment with the effect of sea quarks \cite{10.1093/ptep/ptae060,10.1093/ptep/ptad108}, hyperons semi-leptonic decay \cite{BATRA2014126,Upadhyay2013}, spin distribution \cite{Batra_2014,BATRA201218,doi:10.1142/S0217751X13500620}.
% It provide a stronge base to understand the quark-gluon dynamics. 
% We know that hadrons, in addition to valence quarks, also possess a "dynamic sea" filled with different quark-antiquark ($u\bar u, d\bar d, s\bar s$) flavors and gluons. Several experimental collaborations like STAR [], NuTeV [], FAIR [], $\bar P$ANDA-GSI, BESIII [] has been confirmed the impact of strange and non-strange sea on the static properties of baryons. 
The statistical model assumes the hadrons as an ensemble of quark-gluon Fock states $|\{q^3\},\{i,j,k,l\}\rangle$. Singh $\&$ Upadhyay \cite{JPSingh_2004} developed Fock states of nucleons having speciﬁc color and spin quantum numbers possessing symmetry properties, using statistical notions. The ratio of the magnetic moments of the nucleons, their weak decay constant, and the ratio of SU(3) reduced matrix elements for the axial current is also studied. The main purpose of this paper is to formulate the quadrupole transition moment from spin-$\frac{3}{2}^+$ decuplet to spin-$\frac{1}{2}^+$ octet baryons using the statistical approach. The detailed balance principle is used to calculate the probabilities associated with different Fock states. The individual contribution of strange ($s\bar s$) and non-strange ($u\bar u, d\bar d$) sea quarks is calculated. We investigate the SU(3) flavor symmetry and its breaking parameters for both valence and sea quarks to enhance our analysis.\\
% The statistical model has been successfully applied to calculate the strange and non-strange sea contribution to, charge radii and quadrupole moment [], strange mass correction to hyperons semi-leptonic decay [], masses of octet and decuplet baryons [], spin distribution [], octet magnetic moment and their sum rule []. It provide a stronge base to understand the quark-gluon dynamics.\\
The present work proceeds in the following manner. In sec. 2, a concise explanation of transition quadrupole moment and baryonic wavefunction, including sea components is discussed. Sec. 3 presents an introduction to the statistical model coupled with detailed balance principle. The relevance of statistical parameters to the static properties of baryons is also described. In Sec. 4, we present the computed value of quadrupole transition moment, and compared it with updated experimental as well as theoretical data. The conclusion is provided in the Sec. 5.

\section{Tansition quadrupole moment}
Electric quadrupole moment is one of the most prominent properties that characterize the deformation in the geometrical shape of baryons. The finite value of quadrupole moment depicts the picture of charge distribution inside the baryons in  such a way that either the baryon exhibits an oblate shape (cushion-like structure) or prolate shape (rugby-ball) in the fixed body frame. In the multipole expansion, the charge density operator $\rho$ is expressed up to the limit of $q^2$ as:\\
\begin{equation}
    \rho(q) = e- \frac{q^2}{6}r^2_{B}-\frac{q^2}{6}Q_{B}+...
\end{equation}
Here $q^2$ represents the four-momentum transfer of the virtual photon. The second-order (quadrupole) term of the charge density operator $\rho (q)$ relates to the transition quadrupole moment. It is a measure of the deformation of the charge distribution in a baryon during a transition between two states, particularly from an excited state to ground state.  
% The quadrupole transition moment of baryons ($B^* \rightarrow B + \gamma$) measured the change in the distribution of charge during a transition between two states. The low-lying baryon decuplet to octet transition quadrupole moment
In this section, we calculate the transition quadrupole moments for the radiative decays $B^* \rightarrow B + \gamma$, where $B^*$ and $B$ represent the initial and final state of baryons, respectively. 
% Since the E2 transition involves the contribution of quadrupole moments, it can lead to the transition from the spin $\frac{3}{2}^+$ decuplet to the spin $\frac{1}{2}^+$ octet.
The quadrupole moment transition operator given as \cite{SHARMA2013,PhysRevD.65.073017}:
\begin{equation}
 Q^{'}_{B^{*}\rightarrow B}= 3B'\sum_{i\neq j} e_i\sigma_{iz}\sigma_{jz}+ 3C'\sum_{i\neq j\neq k} e_i\sigma_{jz}\sigma_{kz}
    \end{equation}\\
Here, $\sigma_{iz}$ denotes the z-component of the Pauli spin matrix $\sigma_i$ and $e_i$ is the charge of the $i^{th}$ quark where i = (u,d,s). The constants $B^{'}$ and $C^{'}$ parameterize the contribution of color and orbital space \cite{Buchmann2018}. The transition quadrupole moment can be evaluated from the matrix element of the operator when it is applied to the spin-flavor octet and decuplet wavefunction.
\begin{equation}
Q^{'}_{B_{\frac{3}{2}^+\rightarrow\frac{1}{2}^+}}= \langle B^*_{\frac{3}{2}^+}|Q^{'}|B_{\frac{1}{2}^+}\rangle
\end{equation}
Here, B$^*$ and B correspond to the spin-flavor wavefunction of decuplet and octet baryons, respectively. In the quark model approximation, hadrons consists of constituent quarks such as baryons ($qqq$) and mesons ($q\bar q$), arranged in suitable spin-flavor and color singlet configurations. The valence quark wavefunction of baryon \cite{PhysRevD.49.2211} is written as:\\
\begin{equation}
    \Psi= \Phi (|\phi_{flavor}\rangle .|\chi_{spin}\rangle. |\psi_{color}\rangle.|\xi_{space}\rangle)
\end{equation}
Due to color singlet state, the color wavefunction $|\psi\rangle$ of baryons is entirely anti-symmetric. The valence quarks are presumed to be in S-wave state, implies the symmetric spatial wavefunction $|\xi\rangle$ under the interchange of any two quarks. Therefore, the flavor-spin wavefunction ( $|\phi\rangle$, $|\chi\rangle$) must exhibit complete symmetry to ensure the anti-symmetrization of baryonic system. Additionally, the presence of quark-gluon interaction suggests that baryons are composed not only of valence quarks, but also comprising a ”dynamic sea” having infinite no. of virtual quark-antiquark ($q\bar q$) pairs and gluons. The ”sea” is expressed by flavor, spin, and color quantum numbers. For simplicity, we assumed the sea to be flavorless and the possibilities in spin and color space is written as:\\

\makebox[\textwidth]{ \textbf{Spin}: gg: $ 1\bigotimes1 = 0_s\bigoplus 1_a\bigoplus 2_S$}\\
\makebox[\textwidth]{$q\bar qq\bar q : (\frac{1}{2}\bigotimes \frac{1}{2})\bigotimes(\frac{1}{2}\bigotimes \frac{1}{2}) = (0_a\bigoplus1_s)\bigotimes(0_a\bigoplus1_s)$}\\
\makebox[\textwidth]{\hspace{2cm} = $2(0_s)\bigoplus 1_s\bigoplus2(1_a)\bigoplus2_s$}\\
%\makebox[\textwidth]{\textbf{Color}:  qqq: $3 \bigotimes3 \bigotimes3 = 1_a\bigoplus 8_{ms}\bigoplus 8_{ma}\bigoplus10_s$}\\

\makebox[\textwidth]{\textbf{Color}: gg: $8\bigotimes 8 = 1_s\bigoplus 8_s\bigoplus 8_a\bigoplus 10_a\bigoplus \bar {10_a}\bigoplus 27_s$}\\
\makebox[\textwidth]{$q\bar qq\bar q : (3\bigotimes \bar 3)\bigotimes(3\bigotimes \bar 3) = (1_a\bigoplus8_s)\bigotimes(1_a\bigoplus8_s)$}\\
\makebox[\textwidth]{\hspace{2cm} = $2(1_s)\bigoplus2(8_s)\bigoplus2(8_a)\bigoplus10_s\bigoplus \bar {10_s} \bigoplus 27_s$}\\

The subscript $\textbf{s}$ and $\textbf{a}$ stands for the symmetric and anti-symmetric combinations of
the states, respectively. The Sea spin and color wavefunctions denoted by $H_{0,1,2}$ and $G_{1,8,\bar{10}}$ respectively.
% which fullfill the condition: {$<H_i|H_j> = \delta_{ij}$, $<G_k|G_l> = \delta_{kl}$}.
The total spin-flavor-color wavefunction consists of valence quarks and sea can be written as:\\

\textbf{For octet baryons} \cite{PhysRevD.49.2211}:\\
\begin{equation}
\begin{aligned} |\Phi_{1/2}^{(\uparrow)}> = & \frac{1}{N} [{\Phi_{1}}^{(\frac{1}{2} \uparrow)}H_{0}G_{1} + a_{8} ({\Phi_{8}}^{(\frac{1}{2})}\otimes H_0)^{\uparrow}G_8 + a_{10} {\Phi_{10}}^{(\frac{1}{2}\uparrow)}H_{0}G_{\bar {10}} +\\& b_1({\Phi_1}^{(\frac{1}{2})} \otimes H_1)^{\uparrow} G_1+b_8 ({\Phi_8}^{(\frac{1}{2})} \otimes H_1) ^{\uparrow} G_8 + b_{10} ({\Phi_{10}}^{(\frac{1}{2})} \otimes H_1)^{\uparrow} G_{\bar {10}}\\& + c_8 ({\Phi_8}^{(\frac{3}{2})} \otimes H_1)^{\uparrow} G_8 + d_8 ({\Phi_8}^{(\frac{3}{2})} \otimes H_2)^{\uparrow} G_8\end{aligned}\end{equation}\\
where $N^{2} = 1 + a_{8}^2+a_{10}^2+ b_{1}^{2} +b_{8}^{2} + b_{10}^2+ c_{8}^{2} + d_{8}^{2}$\\

\textbf{For decuplet baryons}\cite{Kaur20167}:\\
\begin{equation}
    \begin{aligned}
  |\Phi_{3/2}^{(\uparrow)}> =& \frac{1}{N'} [a'_0{\Phi_{1}}^{(\frac{3}{2} \uparrow)}H_{0}G_{1} + b'_1({\Phi_1}^{(\frac{3}{2})} \otimes H_1)^{\uparrow} G_1 + b'_8 ({\Phi_8}^{(\frac{1}{2})} \otimes H_1) ^{\uparrow} G_8 +\\&d'_1 ({\Phi_1}^{(\frac{3}{2})} \otimes H_2)^{\uparrow} G_1 + d'_8 ({\Phi_8}^{(\frac{1}{2})} \otimes H_2)^{\uparrow} G_8]   
    \end{aligned}
\end{equation}
where $N^{'2} = a_{0}^{'2} +b_{1}^{'2} +b_{8}^{'2} + d_{1}^{'2} + d_{8}^{'2}$\\
Here, $N$ and $N'$ represent the normalization constant. In the octet wavefunction (eq. 5), the first term $\Phi_{1}^{(\frac{1}{2} \uparrow)}H_{0}G_{1}$ includes the valence part $\Phi_{1}^{(\frac{1}{2} \uparrow)} = \Phi (8,\frac{1}{2},1_c)$ having spin-$\frac{1}{2}$, color singlet and flavor octet. It is combined with sea spin-0 ($H_0$) and color singlet ($G_1$). Similarly, all other terms 
 $\Phi^{(\frac{1}{2})\uparrow}_{b1}$, $\Phi^{(\frac{1}{2})\uparrow}_{b8}$, $\Phi^{(\frac{3}{2})\uparrow}_{d1}$, $\Phi^{(\frac{1}{2})\uparrow}_{c8}$, $\Phi^{(\frac{1}{2})\uparrow}_{d8}$ are written by taking coupling between spins of sea and valence part with suitable C.G. coefficients.
 % securing the overall anti-symmterisation with flavor-8, color-1 and spin-$\frac{1}{2}$ (flavor-10, color-1 and spin-$\frac{3}{2}$).
 The sea having spin 0, 1, and 2 specifies scalar, vecto, and tensor sea, respectively. The coefficients ($a'_0, a_8, a_{10}, b_1, b'_1, b_8, b_{10}, c_8, d'_1, d_8$) in eq. (2) and (3) are the statistical parameters that account for the sea's contribution in terms of scalar, vector, and tensor. In the octet wave function, the coefficients $a_8, a_{10}$ come from the scalar sea, whereas the coefficients $b_1 b_8, b_{10}, c_8$ and $d_8$ are from the vector and tensor sea, respectively. Similarly, in case of decuplet, the coefficient $a'_0$ is associated with scalar sea, and the coefficients $b'_{1}, b'_{8}$ and $d'_1, d'_8$ are associated with vector and tensor sea, respectively. These statistical parameters play a central role in determining the impact of sea quarks and gluons on different low-energy properties of baryons such as masses, semi-leptonic decays, magnetic moment etc. The comprehensive detail regarding the octet and decuplet wavefunctions can be found in refs. \cite{Kaur20167,BATRA201218,PhysRevD.49.2211}.\\
The transition from spin $\frac{3}{2}^+\rightarrow \frac{1}{2}^+$, flavor decuplet to octet ($B_{10}\rightarrow B_8$)  and the overall color singlet state could be possible through 10 different combinations of decuplet and octet wavefunction. The combinations are given below:\\
\begin{center}
    $b'_1(\Phi_1^{(\frac{3}{2})}\bigotimes H_1)^{\uparrow}G_1 \rightarrow \Phi_1^{(\frac{1}{2}\uparrow)}H_0G_1$\\\vspace{2mm}
    {$b'_1(\Phi_1^{(\frac{3}{2})}\bigotimes H_1)^{\uparrow}G_1 \rightarrow \Phi_1^{(\frac{1}{2}\uparrow)}H_0G_1$\\\vspace{2mm}
$b'_1(\Phi_1^{(\frac{3}{2})}\bigotimes H_1)^{\uparrow}G_1 \rightarrow b_1(\Phi_1^{(\frac{1}{2})}\bigotimes H_1)^{\uparrow}G_1$\\\vspace{2mm}
$b'_8(\Phi_8^{(\frac{1}{2})}\bigotimes H_1)^{\uparrow}G_8 \rightarrow a_8\Phi_{8}^{(\frac{1}{2}\uparrow)}H_0G_8$\\\vspace{2mm}
$b'_8(\Phi_8^{(\frac{1}{2})}\bigotimes H_1)^{\uparrow}G_8 \rightarrow b_8(\Phi_8^{(\frac{1}{2})}\bigotimes H_1)^{\uparrow}G_8$\\\vspace{2mm}
$b'_8(\Phi_8^{(\frac{1}{2})}\bigotimes H_1)^{\uparrow}G_8 \rightarrow c_8(\Phi_8^{(\frac{3}{2})}\bigotimes H_1)^{\uparrow}G_8$\\\vspace{2mm}
$b'_8(\Phi_8^{(\frac{1}{2})}\bigotimes H_1)^{\uparrow}G_8 \rightarrow d_8(\Phi_8^{(\frac{3}{2})}\bigotimes H_2)^{\uparrow}G_8$\\\vspace{2mm}
$d'_1(\Phi_1^{(\frac{3}{2})}\bigotimes H_2)^{\uparrow}G_1 \rightarrow \Phi_1^{(\frac{1}{2}\uparrow)}H_0G_1$\\\vspace{2mm}
$d'_8(\Phi_8^{(\frac{1}{2})}\bigotimes H_2)^{\uparrow}G_8 \rightarrow a_8\Phi_{8}^{(\frac{1}{2}\uparrow)}H_0G_8$\\\vspace{2mm}
$d'_8(\Phi_8^{(\frac{1}{2})}\bigotimes H_2)^{\uparrow}G_8 \rightarrow b_8(\Phi_8^{(\frac{1}{2})}\bigotimes H_1)^{\uparrow}G_8$}\\\vspace{2mm}
\end{center}
It is important to mention that we are excluded the wavefucntion terms $a_{10} {\Phi_{10}}^{(\frac{1}{2}\uparrow)}H_{0}G_{\bar {10}}$, $b_{10} ({\Phi_{10}}^{(\frac{1}{2})} \otimes H_1)^{\uparrow} G_{\bar {10}}$ mentioned in eq. 5 as they do not meet the quantum number conditions from decuplet to octet transition. The transition quadrupole moment operator is applied on suitable combinations as:\\
% \begin{equation}
% Q^{'}(B_{\frac{3}{2}^+\rightarrow\frac{1}{2}^+})= \langle B^*_{\frac{3}{2}^+}|Q^{'}|B_{\frac{1}{2}^+}\rangle
% \end{equation}
% The transition quadrupole moment operator mention in eq.2 is applied on the above-mentioned combinations of the $J^P = \frac{3}{2}^+$ decuplet and $J^P = \frac{1}{2}^+$ octet baryons wavefunction as:\\
\begin{equation}
    \begin{aligned}
       \langle\Phi_{3/2}^{(\uparrow)}|{{Q'}}| \Phi_{1/2}^{(\uparrow)}\rangle =& \frac{1}{NN'} [{b'_1b_1}{\langle \Phi_{1}}^{(\frac{3}{2} \uparrow)}|\widehat{Q'}|{\Phi_{1}}^{(\frac{1}{2}\uparrow)}\rangle + {b'_8a_8} \langle{\Phi_8}^{(\frac{3}{2}\uparrow)}|\widehat{Q'}|{\Phi_8}^{(\frac{1}{2}\uparrow)}\rangle + \\& {b'_8 b_8} \langle{\Phi_{8}}^{(\frac{3}{2}\uparrow)}|\widehat{Q'}|{\Phi_{8}}^{(\frac{1}{2}\uparrow)}\rangle + {d'_8 a_8}\langle{\Phi_8}^{(\frac{3}{2}\uparrow)}|\widehat{Q'}|{\Phi_8}^{(\frac{1}{2}\uparrow)}\rangle + ...........
    \end{aligned}
\end{equation}
Certain expression are obtained in terms of statistical parameters $a_0, a_8, b_1, b_8, c_8, d_1, d_8, b'_{1}, b'_{8}, d'_{1}$ and the constants $B^{'}$ and $C^{'}$ as:\\
\begin{equation}
\begin{aligned}
     \boldsymbol{{\Delta^{+}}\rightarrow p} =& b_{8}' a_{8}(\frac{1}{2\sqrt{30}}(B'-5C')) + b_{8}' b_{8}(\frac{1}{\sqrt{360}}(B'-5C')) +b_{8}'c_{8}(\frac{1}{3\sqrt{10}}(B'-5C')) +b_{8}'d_{8}(\frac{-1}{\sqrt{150}}(B'-5C')) -\\& d_{8}'a_{8}(\frac{-1}{2\sqrt{90}}+\frac{1}{2\sqrt{30}}+\frac{1}{\sqrt{90}}(B'-5C'))+d_{8}'b_{8} (\frac{1}{2\sqrt{270}}-\frac{1}{2\sqrt{90}}-\frac{1}{\sqrt{270}}+\frac{1}{\sqrt{540}}-\frac{1}{\sqrt{180}}-\frac{2}{\sqrt{540}})\\&(B'-5C'))
\end{aligned}
\end{equation}
\begin{equation}
     \begin{aligned}
          \boldsymbol{{\Sigma^{*+}}\rightarrow \Sigma^{+}} =& b_{8}'a_{8}(\frac{1}{4\sqrt{30}}(2B'-10C')) +b_{8}'b_{8} (\frac{1}{2\sqrt{360}}(2B'-10C'))+b_{8}'c_{8}(\frac{-1}{6\sqrt{10}}(2B'-10 C'))+\\&b_{8}'d_{8}(\frac{1}{2\sqrt{150}}(2B'-10C'))+d_{8}'a_{8}(\frac{1}{4\sqrt{90}}+\frac{1}{4\sqrt{30}})(2B'-10C')\\&+d_{8}'b_{8} (\frac{-1}{4\sqrt{270}}-\frac{1}{4\sqrt{90}}+\frac{1}{2\sqrt{540}}-\frac{1}{2\sqrt{180}})(2B'-10C')
     \end{aligned}
     \end{equation}
     
\begin{equation}
     \begin{aligned}
        \boldsymbol{{\Sigma^{*-}}\rightarrow \Sigma^{-}} =& b_{8}'a_{8} (\frac{1}{4\sqrt{30}}(2B'+ 2C'))+b_{8}'b_{8} (\frac{-1}{2\sqrt{360}}(2B'+ 2C'))+b_{8}'c_{8}(\frac{-1}{6\sqrt{10}}(2B'+2C'))\\&+b_{8}'d_{8} (\frac{1}{2\sqrt{150}}(2B'+ 2C'))+d_{8}'a_{8}[(\frac{1}{4\sqrt{90}}+\frac{1}{4\sqrt{30}})(2B'+ 2C')+\\&d_{8}'b_{8}(\frac{-1}{12\sqrt{270}}(6B'+ 6C')-(\frac{1}{4\sqrt{90}}-\frac{1}{2\sqrt{540}}-\frac{1}{2\sqrt{180}})(2B'+ 2C'))
     \end{aligned}        
     \end{equation}
     
     \begin{equation}
\begin{aligned}
     \boldsymbol{{\Sigma^{*0}}\rightarrow \Sigma^{0}} =&  b_{1}'a_{0} (\frac{1}{36\sqrt{5}}(6B'-12C'))-b_{1}'b_{1} (\frac{\sqrt{2}}{36\sqrt{15}}(6B'-12C'))+b_{8}'a_{8}(\frac{1}{12\sqrt{30}}(6B'-12C'))\\&- b_{8}'b_{8} (\frac{1}{12\sqrt{360}}(-6B'+ 12C'))-b_{8}'c_{8}(\frac{1}{36\sqrt{10}}(6B'-12C'))+b_{8}'d_{8}(\frac{1}{36}\sqrt{\frac{3}{50}}(6B'-12C'))+\\&d_{1}'a_{0}(\frac{1}{36\sqrt{5}}(6B'-12C'))+ d_{8}'a_{8}(\frac{1}{24\sqrt{90}}(-6B'+ 12C')+\frac{1}{8\sqrt{30}}(-12B'+12C'))\\&+d_{8}'b_{8} (\frac{1}{24\sqrt{270}}-\frac{1}{12\sqrt{540}})(-6B'+ 12C')-(\frac{1}{8\sqrt{90}}-\frac{1}{4\sqrt{180}})(-12B'+12C')) 
\end{aligned}
\end{equation}

 \begin{equation}
       \begin{aligned}
           \boldsymbol{{\Xi^{*0}}\rightarrow \Xi^{0}}=& b_{1}'a_{0} (\frac{C'}{9\sqrt{5}}+b_{1}'b_{1} (\frac{C'}{9}\sqrt{\frac{2}{15}})-b_{8}'a_{8}(\frac{1}{4\sqrt{30}}(-4B'+ 8C'))+b_{8}'b_{8}(\frac{1}{2\sqrt{360}}(-4B' + 8C'))+\\&b_{8}'c_{8}(\frac{1}{6\sqrt{10}}(-4B' + 8C')+\frac{1}{\sqrt{30}}(2B'+ 2C'))+b_{8}'d_{8}(\frac{-1}{6}\sqrt{\frac{3}{50}}(-4B' +8C')+\\&\frac{1}{\sqrt{30}}(4B'+4C'))+d_{1}'a_{0} (\frac{C'}{9\sqrt{5}})+d_{8}'a_{8}(\frac{1}{2\sqrt{90}}+\frac{1}{4\sqrt{30}}-\frac{1}{4\sqrt{90}})(-4B'+ 8C'))+\\&d_{8}'b_{8}(\frac{1}{12\sqrt{270}}(-12B' +24C')+(-\frac{1}{4\sqrt{90}}-\frac{1}{2\sqrt{270}}+\frac{1}{2\sqrt{540}}-\frac{1}{2\sqrt{180}}-\frac{1}{\sqrt{540}})(-4B'+ 8C'))
       \end{aligned}      
       \end{equation}
       
 \begin{equation}     
       \begin{aligned}
           \boldsymbol{{\Xi^{*-}}\rightarrow \Xi^{-} }=& b_{8}'a_{8}(\frac{1}{12\sqrt{30}}(6B' +6C'))+b_{8}'b_{8} (\frac{-1}{6\sqrt{360}}(6B' +6C'))+b_{8}'c_{8}(\frac{-1}{18\sqrt{10}}(6B' +6C'))\\&+b_{8}'d_{8} (\frac{1}{18}\sqrt{\frac{3}{50}}(6B' +6C'))+d_{8}'a_{8}(\frac{1}{12\sqrt{90}}(6B' +3C'(-2+4r))+\frac{1}{4\sqrt{30}}(2B'+2C'))\\&+d_{8}'b_{8} (\frac{-1}{12\sqrt{270}}-\frac{1}{6\sqrt{540}})(6B' +6C')+(\frac{-1}{4\sqrt{90}}-\frac{1}{2\sqrt{180}})(2B'+ 2C'))  
       \end{aligned}
       \end{equation}

\begin{equation}
    \begin{aligned}
       \boldsymbol{{\Sigma^{*0}}\rightarrow \Lambda}  = &   b_{1}'b_{1} (\frac{1}{\sqrt{90}}(3B'-6C'))+b_{1}'a_{0} (\frac{-1}{2\sqrt{15}}(3B'-6C')+ b_{8}'a_{8}(\frac{1}{8\sqrt{10}}(6B'-12C'))\\&+b_{8}'b_{8} (\frac{-1}{4\sqrt{120}}(6B'-12C'))+b_{8}'c_{8}(\frac{1}{2\sqrt{30}}(3B'-6C'))+b_{8}'d_{8}(\frac{1}{2\sqrt{50}}(-3B'+6C'))+\\& d_{1}'a_{0}(\frac{1}{2\sqrt{15}}(-3B'+6C'))+d_{8}'a_{8} (\frac{1}{2\sqrt{90}}(-3B'+3C')+\frac{1}{8\sqrt{30}}(6B'-12C'))+\\&d_{8}'b_{8}(\frac{1}{2\sqrt{270}}(3B'-3C')+\frac{1}{4\sqrt{90}}(3B'-6C')+\frac{1}{2\sqrt{540}}(3B'-3C')+\frac{1}{2\sqrt{180}}(3B'-6C')) 
    \end{aligned}
     \end{equation}
The parameters $B^{'}$ and $C^{'}$ are calculated previously in our article Ref. \cite{{10.1093/ptep/ptae060,10.1093/ptep/ptad108}} using experimental inputs on quadrupole moment.
% The SU(3) flavor symmtery breaking in valence quarks is studied by introducing a mass dependent parameter 'r' in the operator. 
The statistical coefficients possess the greatest significance in themselves as they are linked with the probability distribution of each Fock state in terms of flavor, spin
and color sub-space. To define the set of probabilities for quadrupole transition, statistical approach along with the detailed balance principle is employed, discussed in the subsequent section.
\section{Statistical model and principle of detailed balance}
Zhang et al. \cite{PhysRevD.65.114005} proposed the concept of detailed balance, which states that each physical hadron state can be described as a composite of multiple quark-gluon Fock states. Each Fock state comprises an infinite no. of $q\bar q$ pairs mediated by gluons and expressed as: \\
\begin{equation}
|h \rangle = \sum_{i,j,k,l} C_{i,j,k,l} | \{q^3\},\{i,j,k,l\}\rangle
\end{equation}
Here $\{q^3\}$ represent the valence quarks, $i$, $j$, and $l$ denotes the no. of $\bar q q$ pairs, i.e. $u\bar{u}$, $d\bar{d}$, $s \bar s$ pairs respectively and $k$ is the no. of gluons. 
% Basically, the quarks and gluons present in the Fock states are the intrinsic partons which are multiconnected non-perturbatively to the valence quarks \cite{55}. It is important to mention that a confined gluon within the sea can be categorized into TE (transverse electric) mode with $J^{PC}$ = 1$^{+-}$ and the TM (transverse magnetic) modes with $J^{PC}$ = 1$^{--}$. To keep the parity of the system positive, Fock states with a single gluon considered to be consisting of a TE mode. Similarly, Fock states with $q\bar q$ pairs required to be in p-wave state for maintaining positive parity of the system.  
The probability of finding the baryon in quark-gluon Fock state $|\{q^3)\},\{i,j,k,l\}\rangle$ is $ \rho_{i,j,k,l} = |C_{i,j,k,l}|^2 $ where $\rho_{i,j,k,l}$ satisfy the normalization condition $\sum_{i,j,k,l} \rho_{i,j,k,l} = 1$.\\
% This principle considered that any two neighboring quark-gluon Fock states should be in equilibirium with each other \cite{56}. 
This principle demands that the probability of having baryon in a particular Fock state should be same over the time and it is expressed as: \\
$$\rho_{i,j,k,l} |\{q^{3}\},\{i,j,k,l\}\rangle \rightleftharpoons \rho_{i',j',k',l'}|\{q^{3}\},\{i',j',k',l'\}\rangle$$\\
The transfer between two Fock states occurs in two ways: the go-out rate, which counts the possible split of partons and the come-in rate, which represents the possible recombination of partons. The detailed balance model was used to examine the statistical effects of the nucleon and d–u asymmetry. The model yielded $\bar d - \bar u$ = 0.124, which was consistent with the predictions of the E866/NuSea result of 0.118$\pm$0.012 \cite{PhysRevD.64.052002}. As we mentioned, the effect of sea quarks and gluons is investigated in the form of statistical parameters ($a_0, a_8, b_1, b_8, c_8, d'_{1}, d'_{8}$). The first step is to calculate the probabilities of various Fock states in flavor space using the principle of detailed balance. The transition sub-processes like g $\rightleftharpoons  q \bar q$, g $\rightleftharpoons $ gg, and q $\rightleftharpoons $ qg are used to calculate the probabilities. The probability of having a proton in several Fock states is discussed in Ref \cite{ZHANG2001260}.\\
 % \begin{table*}[h!]{\normalsize\renewcommand{\arraystretch}{1.5}\tabcolsep 4mm \centering\small{ \begin{tabular}{|c|c|c|>{\centering\arraybackslash}p{0.15\linewidth}|>{\centering\arraybackslash}p{0.13\linewidth}|>{\centering\arraybackslash}p{0.13\linewidth}|>{\centering\arraybackslash}p{0.13\linewidth}|} \hline i  & j & $\rho_{i,j,k}$&$\rho_{i,j,0}$&$\rho_{i,j,1}$&$\rho_{i,j,2}$&$\rho_{i,j,3}$\\ \hline 0 &0 &$uud$&0.15888&0.15888&0.09079&0.03782\\  0&1 &$uudd\bar d $&0.07944&0.07944&0.04333&0.01685\\ 1&0&$uudu\bar u $&0.05296&0.05296&0.02888&0.01123\\1&1&$uudu\bar u d\bar d$&0.02648&0.02648&0.01412&0.00529\\   0&2&$uudd\bar d d\bar d$&0.01324&0.01324&0.00706&0.00264\\ 2&0&$uudu\bar u u\bar u$&0.00662&0.00620&0.00353&0.00132\\1&2&$uudu\bar u d\bar d d\bar d$&0.00441&0.00441&0.00232&0.00085\\2&1&$uudu\bar u u\bar u d\bar d$&0.00331&0.00331&0.00174&0.00063\\2&2&$uudu\bar u u\bar u d\bar d d\bar d$&0.00055&0.00055&0.00028&0.00010\\  0&3&$uudd\bar d d\bar d d\bar d $&0.00110&0.00110&0.00058&0.00021\\3&0&$uudu\bar u u\bar u u\bar u$&0.00044&0.00044&0.00023&0.00008\\1&3&$uud u\bar u d\bar d d\bar d d\bar d$&0.00036&0.00036&0.00019&0.00006\\\hline\end{tabular}} \caption{The probabilities of finding the quark–gluon Fock states of the proton
 % } \label{tab:my_label}}\end{table*}
Moreover, when we consider the transition process g $\rightleftharpoons$ $s \bar s$, the probability of all the Fock states modified due to the large mass of strange quark. %It has been known from various experimental and theoretical studies [] that the strange quark plays a unique role in the spectroscopy of baryons. 
This transition can only take place if the gluons have a specific amount of free energy that exceeds the mass of strange quark, i.e. $\varepsilon_g > 2M_s$. The production of $s \bar s$ pairs from gluons is constrained by a suppression factor expressed as $k(1 - C_{l})^{n-1}$ \cite{doi:10.1142/S0217751X03014915}, where n denoted the total no. of partons present in the Fock state. The suppression factor arises from  distribution of free energy of gluons as well as the total energy of partons present in the baryon. The value of $C_{l-1}=\frac{2M_s}{M_B-2(l-1)M_s}$, where $M_B$ is the mass of baryon. The expression for transition rates between gluon and strange $q\bar q$ pairs can be expressed as follows:\\
\begin{equation}
\frac{\rho{i,j,l,k}}{\rho{i,j,l+k,0}}= \frac{k(k-1)(k-2)...1(1-C_0)^{n-2l-1}(1-C_1)^{n-2l}....(1-C_{l-1}^{n+k-2})}{(l+1)(l+2)...(l+k)(l+k+1)}
\end{equation}
The probabilities expression of Fock states are obtained in terms of $\rho_{0000}$ and is different for each baryon. Ref. \cite{PhysRevD.65.114005,doi:10.1142/S0217751X13500620,BATRA2014126} provide the details of the above expression.  
% It is important to note that the value of constraint $(1 - C_{l})^{n-1}$ clearly indicates the difference between double strange baryon and single strange baryon to accommodate $s \bar s$ pairs \cite{58}. 
It has been observed that the involvement of more $\bar s s$ pairs in sea affects the value of $(1 - C_{l})^{n-1}$ and the alters the probabilities of Fock states. In the present paper, we are assuming two $s\bar s$ condensates in sea and calculated the probabilities of each baryon by taking strange quark mass from PDG \cite{Workman:2022ynf}. The Fock states without strange quark content account for the 86$\%$ of Total Fock states while the inclusion of $s \bar s$ reduces this proportion to 80$\%$. The most exciting aspect of having the strange sea through the processes g $\rightleftharpoons$ $ s \bar s$ is that flavor SU(3) symmetry breaking within the sea can be studied. The probabilities of different Fock states after including the $\bar s s$ condensates for $\Sigma^{*0}$ baryon is presented in Table 1. 

\begin{table*}[h!]{\normalsize
\renewcommand{\arraystretch}{1.5}
\tabcolsep 4mm
    \centering
   \small{
    \begin{tabular}{|c|c|c|c|
>{\centering\arraybackslash}p{0.15\linewidth}|>{\centering\arraybackslash}p{0.13\linewidth}|>{\centering\arraybackslash}p{0.13\linewidth}|>{\centering\arraybackslash}p{0.13\linewidth}|} \hline 
       i  & j & l & $\rho_{i,j,l,k}$&k=0&k=1&k=2&k=3\\ \hline
        0 &0 &0&uds&0.12462&0.08065&0.06757&0.05495\\   
  0&0 &1&$s\bar s$&0.06231&0.02667&0.01372&0.02933\\ 
   0  &1 &0&$d\bar d$&0.06231&0.03017&0.01891&0.01151\\ 
    1&0&0&$u\bar u$&0.06231&0.03017&0.01891&0.01151\\ 
    0  &1 &1&$d\bar d s\bar s$&0.03115&0.00949&0.00348&0.00743\\
     1&0&1&$u\bar u s\bar s$&0.03115&0.00949&0.00348&0.00743\\
    1&1&0&$u\bar u d\bar d$&0.03115&0.01128&0.00529&0.00241\\
 1&1&1&$u\bar u d\bar d s\bar s$&0.01557&0.00338&0.00088&0.00188\\   
0&2&0&$d\bar d d\bar d$&0.01038&0.00376&0.00176&0.00080\\ 
    2&0&0&$u\bar u u\bar u$&0.01038&0.00376&0.00176&0.00080\\
    0&0&2&$s\bar ss\bar s$&0.01038&0.00248&0.00404&0.00260\\
    1&2&0&$u\bar u d\bar d d\bar d$&0.00519&0.00140&0.00049&0.00016\\ 
    0&2&1&$d\bar d d\bar d s\bar s$&0.00519&0.00112&0.00029&0.00062\\ 
 1&2&1&$u\bar u d\bar d d\bar d s\bar s$&0.00259&0.00040&0.00007&0.00015\\
2&1&0&$u\bar u u\bar u d\bar d$&0.00519&0.00140&0.00049&0.00016\\
2&0&1&$u\bar u u\bar u s\bar s$&0.00519&0.00112&0.00029&0.00062\\ 
1&0&2&$u\bar u s\bar s s\bar s$&0.00519&0.00082&0.00134&0.00085\\
0&1&2&$d\bar d s\bar s s\bar s$&0.00519&0.00082&0.00134&0.00085\\  
 %1&2&2&$u\bar u d\bar d d\bar d s\bar s s\bar s$&0.00045&0.00001&0.00001&0.000009\\ 
2&0&2&$u\bar u u\bar u s\bar s s\bar s$&0.00086&0.00009&0.00014&0.00009\\ 
0&2&2&$d\bar d d\bar d s\bar s s\bar s$&0.00086&0.00009&0.00014&0.00009\\ 
2&1&1&$u\bar u u\bar u d\bar d s\bar s$&0.00259&0.00040&0.00007&0.00015\\
%2&1&2&$u\bar u u\bar u d\bar d s\bar s s\bar s$&0.00056&0.00001&0.00001&0.00001\\
0&3&0&$d\bar d d\bar d d\bar d$&0.00086&0.00023&0.00008&0.00002\\
3&0&0&$u\bar u u\bar u u\bar u$&0.00086&0.00023&0.00008&0.00002\\\hline
\end{tabular}}
    \caption{The probabilities of each quark–gluon Fock states for $\Sigma^{*0}$, including the process g $\rightleftharpoons$ $s \bar s$}
    \label{tab:my_label}
} 
\end{table*}
The statistical model is now applied to calculate the relative probabilities of various Fock states ($|u \bar ug\rangle$, $|d \bar dg\rangle$, $|s \bar sg\rangle$, $|u \bar u d\bar d\rangle$, $|u \bar u d\bar dg\rangle$ ) in spin and color space. This approach relies on counting the multiplicities in spin and color space for all possible Fock states configurations in both valence and sea sector. Appropriate ratio's are used to defined the set of multiplicities. A general representation of the multiplicities ratio is given as follows:\\
$$\frac{\rho_{j_1 j_2}}{\rho_{j'_1 j'_2}} =\frac{p(x_1)q(x_2)r(x_3)}{p'(x'_1)q'(x'_2)r'(x'_3)}$$
where $p (p'),q(q')$, and $r(r')$ are the corresponding multiplicities. $x_1$ is the probability of valence part having spin $j_1$. Similarly, $x_2$ is the probability of sea part having spin $j_2$. And $x_3$
represents the probability that the total angular momentum is $\frac{1}{2}$ ($\frac{3}{2}$) for octet (decuplet) baryons. 
% multiplicities for all Fock states are computed as $\rho_{p,q}$ where the relative probability associated with ’Spin p’ for the valence part and ’Spin q’ for sea components. Accordingly, the resultant spin should be $\frac{1}{2}$ ($\frac{3}{2}$) for octet (decuplet) baryons. 
In similar fashion, the probabilities are also derived for color space. The statistical coefficients are simply the sum of product of multiplicities, each multiplied by a common factor ("c") and multiplicity factor (n) for each Fock state. We used the flavor probabilities derived from the detailed balancing principle to determine the common parameter "c" for each possible sea combination. There are two reasons to compare multiplicities in Fock states. The first one is to find the value of common multiplier “c” and the second one is to compute the probability of each substate by keeping in mind the specified spin and color quantum numbers. Detailed calculations is discussed in Ref. \cite{Kaur20167,BATRA2014126,JPSingh_2004}. The sum of the total probabilities provides the statistical coefficients $a_0, a_8, b_1, b_8, c_8, d_8, a'_{0}, b'_{1},b'_{8}$ for both decuplet and octet baryons. In Table 2, the total of probabilities of each Fock state for $\Sigma^{*0}$ are presented. For each baryon, the statistical coefficients presented in Tables 3 and 4, including strange (highlighted in bold) and non-strange sea.  The scalar, vector and tensor sea contribution are defined by these coefficients. The statistical coefficients are of paramount importance in elucidating the mechanism of 'sea quarks' that contribute to the static properties of baryons like masses \cite{10.1093/ptep/ptx068}, semi-leptonic decays \cite{Upadhyay2013}, charge radii \cite{10.1093/ptep/ptae060,10.1093/ptep/ptad108} etc.
 \begin{table*}[h!]
{\normalsize
\renewcommand{\arraystretch}{1.5}
\tabcolsep 4.5mm\centering  \small{   
\begin{tabular}{|c|c|c|c|c|c|} \hline
States  & $H_0G_1$&$H_1G_1^{(3/2)}$&$H_1G_8^{(1/2)}$&$H_2G_1^{(3/2)}$&$H_2G_8^{(1/2)}$\\ \hline   
$|gg \rangle$ & 0.01126&0&0.02252&0.01126&0.02252\\ \hline   
$|u\bar ug\rangle$&0.00274&0.00274&0.01097&0.00274&0.01097\\ \hline   
$|d\bar d g \rangle$& 0.00274&0.00274&0.01097&0.00274&0.01097\\ \hline  
$|s\bar s g \rangle$&0.00242&0.00242&0.00969&0.00242&0.00969\\ \hline  
$|u\bar u u\bar u\rangle$&0.00173&0&0.00346&0.00173&0.00346\\ \hline   $|d\bar d d\bar d \rangle$&0.00173&0&0.00346&0.00173&0.00346\\ \hline   
$|u\bar u d\bar d \rangle$&0.00283&0.00283&0.01132&0.00283&0.01132\\ \hline
 $|u\bar u s\bar s\rangle$&0.00283&0.00283&0.01132&0.00283&0.01132\\\hline
 $|d\bar d s\bar s\rangle$& 0.00283& 0.00283& 0.01132& 0.00283&0.01132\\\hline
 $|s\bar s s\bar s\rangle$&0.00173&0&0.00346&0.00173&0.00346\\ \hline   $|u\bar u s\bar  sg\rangle$&0.00020&0.00061&0.00459&0.00041&0.00330\\ \hline  
 $|u\bar u d\bar d g\rangle$&0.00024&0.00073&0.00588&0.00049&0.00392\\ \hline  
 $|d\bar d  s\bar s g \rangle$ &0.00020&0.00061&0.00459&0.00041&0.00330\\ \hline 
 $|d\bar d d\bar d g \rangle$ &0.00010&0.00020&0.00162&0.00020&0.00162\\ \hline  
 $|s\bar s s\bar s g\rangle$&0.00006&0.00013&0.00107&0.00013&0.00107\\ \hline
 $|u\bar u gg \rangle$&0.00051&0.00102&0.00817&0.00102&0.00817\\ \hline  $|d\bar d gg \rangle$&0.00051&0.00102&0.00817&0.00102&0.00817\\ \hline  $|s\bar s gg \rangle$&0.00037&0.00074&0.00593&0.00074&0.00593\\\hline
 $|g gg \rangle$& 0.00785& 0.00785& 0.0157& 0.00785&0.0157\\\hline
 $|u\bar u u\bar u g \rangle$&0.00010&0.00020&0.00162&0.00020&0.00162\\ \hline
 $|0\rangle$&0.12462&&&&\\ \hline   Total  &0.1676&0.0295&0.15583&0.04531&0.15129\\ 
 &($a_0$)&($b_1$)&($b_8$)&($d_1$)&($d_8$)\\\hline 
 \end{tabular}} 
 \caption{Coefficients for $\Sigma^{*0}$}
 \label{tab:my_label}}
 \end{table*}

\begin{table*}[h!]{\normalsize
 \renewcommand{\arraystretch}{1.5}
 \tabcolsep 4.5mm
    \centering
   \small{
     \begin{tabular}{ccccccc} \hline 
        Coefficients  & $a_0$&$a_8$&$b_1$&$b_8$&$c_8$&$d_8$\\ \hline  
        p & 0.188&0.089&0.018&0.105&0.065&0.053\\
        & \textbf{0.179}&\textbf{0.103}&\textbf{0.021}&\textbf{0.130}&\textbf{0.067}&\textbf{0.061}\\
        \hline  
 $\Sigma^+$& 0.141&0.104&0.025&0.015&0.071&0.061\\
 &\textbf{0.157}&\textbf{0.106}&\textbf{0.024}&\textbf{0.142}&\textbf{0.069}&\textbf{0.072}\\
 \hline  
     $\Sigma^-$& 0.141&0.104&0.025&0.015&0.071&0.061\\
     &\textbf{0.156}&\textbf{0.090}&\textbf{0.024}&\textbf{0.122}&\textbf{0.069}&\textbf{0.072}\\
     \hline 
$\Sigma^0(\Lambda)$&0.168&0.101&0.024&0.141&0.081&0.059\\
     &\textbf{0.189}&\textbf{0.094}&\textbf{0.023}&\textbf{0.126}&\textbf{0.058}&\textbf{0.061}\\
        \hline  
$\Xi^-$&0.126&0.105&0.026&0.153&0.077&0.060\\  &\textbf{0.161}&\textbf{0.106}&\textbf{0.024}&\textbf{0.141}&\textbf{0.070}&\textbf{0.071}\\
        \hline  
$\Xi^0$&0.126&0.105&0.026&0.153&0.077&0.060\\
&\textbf{0.161}&\textbf{0.105}&\textbf{0.024}&\textbf{0.138}&\textbf{0.068}&\textbf{0.071}\\
\hline 
 \end{tabular}}
    \caption{Statistical parameters for spin-$\frac{1}{2}$ octet baryons in SU(3) symmetric sea
and its breaking}
    \label{tab:my_label}
}
\end{table*}

\begin{table*}[h!]{\normalsize
 \renewcommand{\arraystretch}{1.5}
 \tabcolsep 4.5mm
    \centering
   \small{
     \begin{tabular}{cccccc} \hline 
        Coefficients  & $a'_{0}$&$b'_{1}$&$b'_{8}$&$d'_{1}$&$d'_{8}$  \\\hline  
    $\Delta^+$&0.200&0.026&0.158&0.044&0.154\\ &\textbf{0.162}&\textbf{0.027}&0\textbf{.149}&\textbf{0.042}&\textbf{0.144}\\
    \hline  
$\Sigma^{*+}$&0.142&0.020&0.139&0.035&0.132\\
&\textbf{0.152}&\textbf{0.029}&\textbf{0.169}&\textbf{0.044}&\textbf{0.160}\\
\hline  
$\Sigma^{*-}$&0.142&0.020&0.139&0.035&0.132\\ &\textbf{0.152}&\textbf{0.028}&\textbf{0.160}&\textbf{0.043}&\textbf{0.154}\\
\hline 
$\Sigma^{*0}$&0.178&0.027&0.166&0.044&0.160\\ &\textbf{0.167}&\textbf{0.029}&\textbf{0.155}&\textbf{0.045}&\textbf{0.151}\\
\hline   
$\Xi^{*-}$&0.137&0.031&0.188&0.045&0.180\\ &\textbf{0.151}&\textbf{0.025}&\textbf{0.146}&\textbf{0.039}&\textbf{0.140}\\
\hline  
$\Xi^{*0}$&0.137&0.031&0.188&0.045&0.180\\ &\textbf{0.155}&\textbf{0.028}&\textbf{0.161}&\textbf{0.043}&\textbf{0.155}\\
\hline   
    \end{tabular}}
    \caption{Statistical parameters for spin-$\frac{3}{2}$ decuplet baryons in SU(3) symmetric sea
and its breaking}
    \label{tab:my_label}
}
\end{table*}
\section{Numerical results and discussion}
We know that electric quadrupole moment is one of the most valuable property that shed light on inner quark-gluon dynamics of baryons. It is related to the distribution of electric charge within the baryons. Similarly, the analysis of the quadrupole transition moment is also useful as it significantly enhances our understanding to the internal configuration and provides details of the deformation in the structure of baryons. The quadrupole transition moment is a higher-order electromagnetic transition as compared to the dipole transition and is characterize by a change in quadrupole moment. We calculated the transition quadrupole moment from decuplet (spin- $\frac{3}{2}^+$) to octet (spin- $\frac{1}{2}^+$) baryons using the statistical approach. The operator mention in eq. 2 is applied to the spin-flavor baryonic wavefunction. The strength of quadrupole transition moment is influenced by several factors such as spin, parity, angular momentum, etc. The key interest of this work is to understand the impact of 'sea' to the decuplet $\rightarrow$ octet baryons transitions. Throughout this work, two kinds of parameters are used i.e. one is the statistical coefficients ($b_1,b_8,c_8,d_8,,b'_1,b'_8,d'_1,d'_8$) and the other one belongs to the operator (B$'$ and C$'$) which specify the color and spatial contribution of valence quarks. 
%The value of B$'$ and C$'$ is calculated in our previous articles [], fitted from the experimental inputs. 
In order to determine the individual contribution of sea components, the statistical parameters categorized in terms of spin-0 (scalar sea), spin-1 (vector sea) and spin-2 (tensor sea). The statistical coefficients related to each specific sea are taken as non-zero while the other coefficients are set to zero. For example- In case of scalar sea contribution, assuming $b_{1,8,10}, c_8, d_8$ =0 to be zero, the coefficients $a_{0,8,10}, d_8$ =0
for vector sea and for tensor sea the coefficients taken as $a_{0,8,10}, b_{1,8,10}, c_8$ =0 for octet
baryons. The statistical framework includes different models such as Model C, P and D that allows us to investigate the impact of sea dynamics on certain properties of baryons. Model C is the stripped down model which contains various quark-gluon Fock states and assumes the equally likelihood of each Fock state. We present our numerical results for the transition quadrupole moment of baryons including the interaction of sea quark-gluon in Table 7. The impact of SU(3) symmetry breaking (in both sea and valence quarks) to the transition quadrupole moment is also studied. Few points are discussed below that need to be addressed.
\begin{itemize}
    \item $\textbf{Breaking in sea}$\\
In the context of SU(3) flavor symmetry, the concept of flavor breaking in sea is driven by the mass difference between up(u),down(d) and strange(s) quark. Since the mass of the strange quark is significantly higher, the formation of strange $q\bar q$ pairs via gluons is suppressed. To quantify this suppression, a factor $(1-C_l)^{n-1}$ is introduced which influences the probability of various Fock states containing strange quark condensates. Needless to say, the value of  $(1 - C_{l})^{n-1}$ clearly describe the distinction between double strange baryon and single strange baryon to facilitate $s \bar s$ pairs \cite{Kaur20166}. The increase in $\bar s s$ pairs within doubly strange particles, leads to a decrease in the value of suppression factor and affects the overall probability of Fock states. The shift in the probabilities directly modify the statistical parameters; this further changes the transition quadrupole moment likewise. For the case of $\Sigma^{*-}\rightarrow \Sigma^-$, $\Xi^{*0}\rightarrow \Xi^0$, $\Xi^{*-}\rightarrow \Xi^-$, $\Sigma^{*0}\rightarrow \Lambda$ transition, a decrease in the computed value is observed in Table 5. This signifies that the transition quadrupole moment of decuplet $\rightarrow$ octet baryons linked to the probabilities involving the presence of $s\bar s$ pairs. The value of transition moment deviated upto 60$\%$ after assuming the strange sea, as shown in Table 5. The maximum deviation is noted for the $\Sigma^{*+}\rightarrow \Sigma^+$ transition and the deviation is negligible for the case of  $\Sigma^{*0}\rightarrow \Sigma^0$ transition.
\item $\textbf{Breaking in valence}$\\
To enlighten the breaking in valence, we used a mass-correction parameter 'r' that directly in-corporates the mass of strange quark in the relevant operator. It is
defined as r =$\frac{\mu_s}{\mu_d}$ \cite{PhysRevD.49.2211} where $\mu_s$ and $\mu_d$ are the magnetic moments of the strange and down quark, respectively. The best-fit value of ’r’ is calculated in our previous article \cite{10.1093/ptep/ptae060} (r = 0.850). It is quite interesting to observe that results get deviated upto 20$\%$ after applying the strange mass correction in valence. In $\Delta^+\rightarrow p$ transition, the quark content is $uud$, so the transition moment value remains the same as no strange quark is present. Also, for the case of $\Sigma^{*0}\rightarrow \Lambda$ transition, the terms containing parameter 'r' cancel out with each other and the value remains unchanged.
\item To enhance the vision of sea dynamics, the individual contribution of octet and decuplet sea is calculated shown in Tables 6 and 7. Also, the participation of scalar+tensor,  vector+tensor and scalar+vector sea is calculated for both octet and decuplet sector. We observed dominance of the octet vector+tensor sea and it is contributed upto 70$\%$ to the transition quadrupole moment. The scalar+tensor and scalar+vector sea contribution is less comparable but it cannot be neglected. On the other side, the contribution of spin-0 (scalar) of decuplet sea is totally zero, shown in Table 7. As the coupling of constituent quarks with spin-0, 1, and 2 (scalar, vector and tensor sea respectively) leads to a significant no. of possible spin states. Therefore, it might be possible that the chances of mediating transition from decuplet $\rightarrow$ octet is more with higher spin (spin- 1, 2) as compared to scalar sea (spin-0). 
\end{itemize}
\begin{table*}[h!]{\normalsize
\renewcommand{\arraystretch}{1.0}
\tabcolsep 0.1cm
    \centering
   \small{
   \begin{tabular}{ccccccccc}\toprule\hline
           &           & \multicolumn{4}{c}{Statistical model} &  &\\\cmidrule{3-5}
   Baryons & \Longunderstack{Quadrupole transition\\ moment}&\Longunderstack{SU(3)\\ Symmetry \\($g\rightarrow u\bar u, d\bar d$)}& \Longunderstack{SU(3) breaking \\in sea\\($g\rightarrow u\bar u, d\bar d, s\bar s$)}& \Longunderstack{SU(3) breaking\\ in \\ sea+valence} %&   \Longunderstack{Vector\\+Scalar\\Sea}& \Longunderstack{Scalar\\+Tensor\\Sea}& \Longunderstack{Tensor\\+Vector\\ Sea} 
   &\Longunderstack{Without\\sea}\\\midrule%&\Longunderstack{Ref.[]}&\Longunderstack{Ref.[]}\\
   \toprule \vspace{2mm}
    $\Delta^+ \rightarrow p$  &-0.2280B$'$ + 1.1403C$'$ &-0.1058&-0.1576&-0.1576&-0.0331\\ \vspace{2mm}
    $\Sigma^{*+}\rightarrow \Sigma^+$  &-0.1583B$'$ + 0.7918C$'$&-0.0012&-0.0101&-0.0116&0.0204\\ \vspace{2mm}
    $\Sigma^{*-}\rightarrow \Sigma^-$ &-0.5265B$'$ - 0.5265C$'$&0.0526&0.0373&0.0422&0.0047\\ \vspace{2mm}
    $\Sigma^{*0}\rightarrow \Sigma^0$ &0.6367B$'$ - 0.5866C$'$&0.0513&0.0513&0.0558&0.0079\\ \vspace{2mm}
    $\Xi^{*0}\rightarrow \Xi^0$ &1.4559B$'$ - 1.8239C$'$&0.1627&0.0937&0.0823&-0.0073\\ \vspace{2mm}
    $\Xi^{*-}\rightarrow \Xi^-$ &-0.4654B$'$ - 0.4654C$'$ &0.0465&0.0389&0.0279&0.0041\\ \vspace{2mm}
    $\Sigma^{*0}\rightarrow \Lambda$ &0.4468B$'$ - 1.0467C$'$&0.0957&0.0415&0.0415&-0.0011\\ \hline \bottomrule              \end{tabular}
              }}
         \caption{Quadrupole transition moment from decuplet (spin-$\frac{3}{2}^+$) to octet (spin-$\frac{1}{2}^+$) baryons in the units of [fm$^2$]}
         \label{tab:my_label}
     \end{table*}  
\begin{table*}[h!]{\normalsize
 \renewcommand{\arraystretch}{1.5}
 \tabcolsep 4.5mm
    \centering
   \small{
     \begin{tabular}{ccccccccc} \hline 
        Transitions &scalar sea  &vector sea &tensor sea   &vector+tensor\\
        & (spin-0)&(spin-1)&(spin-2)& sea\\\hline  
$\Delta^+ \rightarrow p$ &0.0367&-0.0817&-0.0338&-0.1155\\\hline      
 $\Sigma^{*+}\rightarrow \Sigma^+$& 0.1646&-0.2118&0.0382&-0.173\\
\hline  
$\Sigma^{*-}\rightarrow \Sigma^-$& -0.0389&0.0704&-0.0108&0.0596\\
 \hline  
$\Sigma^{*0}\rightarrow \Sigma^0$& 0.0111&0.0600&0.0060&0.066\\
\hline 
 $\Xi^{*0}\rightarrow \Xi^0$&-0.0517&0.0292&0.0467&0.0759\\
\hline 
$\Xi^{*-}\rightarrow \Xi^-$&-0.0263&0.0457&-0.0061&0.0396\\
\hline 
$\Sigma^{*0}\rightarrow \Lambda$&-0.3149&0.0934&-0.0347&0.0587\\
\hline 
 \end{tabular}}
    \caption{Octet sea contribution}
    \label{tab:my_label}
}
\end{table*}

\begin{table*}[h!]{\normalsize
 \renewcommand{\arraystretch}{1.5}
 \tabcolsep 4.5mm
    \centering
   \small{
     \begin{tabular}{cccccccc} \hline 
        Transitions &scalar sea  &vector sea &tensor sea   &vector + tensor\\
         & (spin-0)&(spin-1)&(spin-2)& sea\\\hline  
$\Delta^+ \rightarrow p$ &0&-0.0142&-0.0524& -0.0666\\\hline      
 $\Sigma^{*+}\rightarrow \Sigma^+$&0 &0.0131&0.0262&0.0393\\
\hline  
$\Sigma^{*-}\rightarrow \Sigma^-$&0 &-0.0035&0.0131&0.0096\\
 \hline  
$\Sigma^{*0}\rightarrow \Sigma^0$&0&-0.0216&0.0198&-0.0018\\
\hline 
 $\Xi^{*0}\rightarrow \Xi^0$&0&-0.0348&0.0349&0.0001\\
\hline 
$\Xi^{*-}\rightarrow \Xi^-$&0&0.0023&0.0086&0.0109\\
\hline 
$\Sigma^{*0}\rightarrow \Lambda$&0&0.0538&0.0612&0.115\\
\hline 
 \end{tabular}}
    \caption{Decuplet sea contribution}
    \label{tab:my_label}
}
\end{table*}
In general, if the concept of 'sea' is excluded completely from the statistical model, the computed results get deviated upto 97$\%$. The value of the quadrupole transition moment without 'sea' is shown in table 5. The statistical model predicted an oblate shape for $\Delta^+\rightarrow p$ and $\Sigma^{*+}\rightarrow \Sigma^+$ transitions which are found to be consistent with experimental data and other phenomenological approaches \cite{SHARMA2013,doi:10.1142/S0217732395001137, PhysRevD.65.073017, ParticleDataGroup:2024cfk}. Also, for rest of the transitions $\Sigma^{*-}\rightarrow \Sigma^-$, $\Sigma^{*0}\rightarrow \Sigma^0$, $\Xi^{*0}\rightarrow \Xi^0$, $\Xi^{*-}\rightarrow \Xi^-$ and $\Sigma^{*0}\rightarrow \Lambda$, a prolated shape is observed, aligned with ref. \cite{doi:10.1142/S0217732395001137,Li2019}. Table 8 displays a comparison of our computed results with available experimental data and different theoretical models.
 \begin{table*}[h!]{\normalsize
 \renewcommand{\arraystretch}{1.5}
 \tabcolsep 2.5mm
    \centering
   \small{
     \begin{tabular}{cccccccc} \hline 
        Models  &  $\Delta^+ \rightarrow p$& $\Sigma^{*+}\rightarrow \Sigma^+$&   $\Sigma^{*-}\rightarrow \Sigma^-$& $\Sigma^{*0}\rightarrow \Sigma^0$& $\Xi^{*0}\rightarrow \Xi^0$& $\Xi^{*-}\rightarrow\Xi^-$&
        $\Sigma^{*0}\rightarrow \Lambda$\\ \hline  
        Exp. data\cite{ParticleDataGroup:2024cfk} &-0.0846$\pm0.0033$&-&-&-&-&-&\\\hline      
        $\chi$CQM\cite{SHARMA2013} & -0.0846&-0.0864&-0.0018&-0.0441&-0.0864&-0.0018&-0.0733\\
        \hline  
 Skyrme \cite{doi:10.1142/S0217732395001137}& -5.2&-0.93&0.93&0.0&2.91&-2.91&-4.83\\
 \hline  
     GPM\cite{PhysRevD.65.073017}& -0.0082&-0.0076&0.014&-0.031&-0.031&0.007&-0.041\\
     \hline 
HBChPT\cite{Li2019}&-0.0075&0.0046&0.007&-0.019&0.0046&0.007&0.059\\
\hline 
Our&-0.1576&-0.0116&0.0422&0.0558&0.0823&0.0279&0.0415\\
\hline 
 \end{tabular}}
    \caption{Comparison of our results with different theoretical approaches}
    \label{tab:my_label}
}
\end{table*}
\section{Conclusion}
In the present work, we are concerned with the transition quadrupole moment from decuplet ($\frac{3}{2}^+$) $\rightarrow$ octet ($\frac{1}{2}^+$) baryons using a statistical approach. In the statistical model, sea quarks and gluons are treated as intrinsic partons and together with the principle of detailed balance, they illuminate the intrinsic structure of the baryons. The principle of detailed balance is used to determine the probabilities for quark-gluon Fock states in terms of statistical parameters ($a_0, a_8, a_{10}, b{'}_1, b'_{8}, c'_{8}, d'_{8}$). To appreciate the importance of flavor symmetry and its breaking in sea, a suppression factor $(1-C_{l})^{n-1}$ is used that includes strange mass corrections directly to the various Fock states. The individual probabilities of various Fock states allow us to check the impact of various sea configurations on the different static properties of baryons. With the help of statistical parameters, the contribution of scalar, vector, and tensor sea investigated separately. Based on our analysis, a prolate and an oblate shape is observed for $\Sigma^{*-}\rightarrow \Sigma^-$, $\Sigma^{*0}\rightarrow \Sigma^0$, $\Xi^{*0}\rightarrow \Xi^0$, $\Xi^{*-}\rightarrow \Xi^-$, $\Sigma^{*0}\rightarrow \Lambda$ and $\Delta^+\rightarrow p$, $\Sigma^{*+}\rightarrow \Sigma^+$ transitions respectively. The key observation of this work is that the octet sea is more actively involved in the quadrupole transition moment as compared to the decuplet sea. The transition quadrupole moment is affected by the spin contribution, flavor symmetry, and charge distribution, among other factors, etc. The sea's spin contribution is reflected in the outcome as spin-1 (vector sea) + spin-2 (tensor sea) contributed more significantly. We concluded that our findings provide significant insights into baryon structure, which in turn motivates further experiments for detailed analysis. The most appealing aspect is that no additional parameters are required in the statistical framework. Also, it is essential to note that our calculations are performed in a non-relativistic frame with the energy scale of order 1 GeV$^2$.
     % \end{landscape}
% \printbibliography

\end{document}